# Atomically Controlled Tunable Doping in High Performance $WSe_2$ Devices

Chin-Sheng Pang, Terry Y.T. Hung, Ava Khosravi, Rafik Addou, Qingxiao Wang, Moon J. Kim, Robert M. Wallace, and Zhihong Chen*

C.-S. Pang, T.Y.T. Hung, Prof. Z. H. Chen
Birck Nanotechnology Center
Department of Electrical and Computer Engineering
Purdue University
1205 W State St, West Lafayette, IN 47907, USA
Email: zhchen@purdue.edu

A. Khosravi, Dr. R. Addou, Q. Wang, Prof. M. J. Kim, Prof. R. M. Wallace
Department of Materials Science and Engineering
University of Texas at Dallas
800 West Campbell Road, Richardson, TX 75080, USA

Dr. R. Addou
School of Chemical, Biological, and Environmental Engineering
Oregon State University
Corvallis, OR 93771, USA

**Abstract:**

Two-dimensional transitional metal dichalcogenide (TMD) field-effect transistors (FETs) are promising candidates for future electronic applications, owing to their excellent transport properties and potential for ultimate device scaling. However, it is widely acknowledged that substantial contact resistance associated with the contact-TMD interface has impeded device performance to a large extent. It has been discovered that $O_2$ plasma treatment can convert $WSe_2$ into $WO_{3-x}$ and substantially improve contact resistances of p-type $WSe_2$ devices by strong doping induced thinner depletion width. In this paper, we carefully study the temperature dependence of this conversion, demonstrating an oxidation process with a precise monolayer control at room temperature and multilayer conversion at elevated temperatures. Furthermore, the lateral oxidation of $WSe_2$ under the contact revealed by HR-STEM leads to potential unpinning of the metal Fermi level and Schottky barrier lowering, resulting in lower contact resistances. The p-doping effect is attributed to the high electron affinity of the formed $WO_{3-x}$ layer on top of the remaining $WSe_2$ channel, and the doping level is found to be dependent on the $WO_{3-x}$ thickness that is controlled by the temperature. Comprehensive materials and electrical characterizations are presented, with a low contact resistance of ~528 Ω μm and record high on-state current of 320 μA/μm at -1V bias being reported.

## 1. Introduction

Two-dimensional (2D) transitional metal dichalcogenides (TMDs) have attracted wide attention, owing to their excellent material properties and potential applications in post-CMOS,[1–8] neuromorphic computing,[9–11] as well as flexible electronics.[12–15] Studying semiconducting TMDs (e.g. $MoS_2$, $WSe_2$) as the channel material for field-effect transistors (FETs) is one of the most vital research areas, due to their superior transport properties and ultra-thin body thickness for ultimate device scaling.[16,17] However, how to make good contacts remains a big challenge for TMD FETs, while it is widely acknowledged that minimizing contact resistance ($R_C$) is essential to obtain high performance devices and reveal intrinsic TMD properties.[18–22]

In general, there are two strategies to optimize current injection at a metal/semiconductor (MS) interface. One method is to select a metal contact with the preferred work function for electron or hole injection, given no strong Fermi level pinning at the contact interface, which is typically not the case for TMDs.[23–25] The other method is to dope the semiconductor degenerately to reduce the depletion width of the MS junction.[26–30] We will show in our paper that by controlling the temperature at which a multi-layer $WSe_2$ FET device is exposed to direct $O_2$ plasma, the number of top $WSe_2$ layers gets converted into $WO_{3-x}$ ($0 < x < 1$) can be precisely controlled, which in turn determines the p-type doping level in the device. Moreover, this conversion is found to extend into the contact area by tens of nanometers, which can possibly unpin the Fermi level of the metal contact and dope a small segment of $WSe_2$ underneath the contact, resulting in reduced contact resistance of the device. The use of controlled oxidation at the Sc/$WSe_2$ interface has also been recently reported to produce an optimized Schottky junction which can be controlled to exhibit n- or p-type transport.[31]

In our unique $O_2$ plasma treatment, the top few $WSe_2$ layers are converted into $WO_{3-x}$ that behaves as a p-type doping layer for the underlying $WSe_2$ due to its high electron affinity. Different from the previously reported self-limiting oxidation of only the topmost $WSe_2$ layer in a remote plasma environment,[32] a direct $O_2$ plasma is employed in our process. Interestingly, we found that the doping level can be tuned from non-degenerate to degenerate by increasing the treatment temperature which directly controls the number of $WSe_2$ layers that get converted into $WO_{3-x}$. We further demonstrate low contact resistance of 528 Ω μm and a record high hole current in a scaled $WSe_2$ FET, shedding light on a promising path in the quest for high-performance electronics. In addition, the achieved p-type doping on $WSe_2$ has excellent air stability, precise doping level control, and is an industry compatible process.

**2. Results and Discussions**

Two types of Schottky barrier (SB) device structures (2 or 4-probe) were implemented, as shown schematically in Figure 1a, b. A scanning electron microscope (SEM) of one of our 4-probe devices is shown in Figure 1c with dimension being labelled. The critical fabrication processes are described in the Experimental Section. High resolution scanning transmission electron microscope (HR-STEM) image and electron energy loss spectroscopy (EELS) line scan across a multi-layer $WSe_2$ flake underneath of a Ti/Pd contact are shown in Figure 1d. Band diagrams and working principles of SB devices are illustrated in the insets of Figure 1e. It is known that the Fermi level of metal contact is pinned close to the mid-gap of $WSe_2$.[33,34] Therefore, band movements modulated by scanning the back-gated voltage ($V_B$) lead to either hole injection from the source to the valence band at negative $V_B$ or electron injection from the drain to the conduction band at positive $V_B$. Consequently, transfer characteristics of a pristine $WSe_2$ device exhibit a typical ambipolar behavior, as shown in Figure 1e.

## 2.1. Material Analyses

Different samples were exposed to a direct $O_2$ plasma at various conditions. Simply comparing color contrast of the samples before and after exposure under an optical microscope, it was rather clear that process temperature is more effective than plasma power or exposure time in controlling the $WSe_2$ oxidation process. Optical images and details are presented in Section I (Supporting Information). Raman spectra of three sets of pristine bi-layer and tri-layer CVD $WSe_2$ samples are presented in the top panels of Figure 2a. Consistent with previous reports,[35,36] the $^1B_{2g}$ Raman mode at 310 $cm^{-1}$ only appears in multi-layers and bulk $WSe_2$ but does not show up in monolayers. In the bottom left panel of Figure 2a, the $^1B_{2g}$ peak vanished after the room temperature (RT) $O_2$ plasma treatment on all three bi-layer samples, indicating that only the bottom $WSe_2$ layer was left while the top layer was oxidized. In contrast, the 310 $cm^{-1}$ peak remained in the spectra for all three tri-layer samples (right bottom panel), suggesting that only the top layer was converted to oxide while the bottom two layers were intact after the RT treatment. Interestingly, when the temperature was raised to 150 $^oC$, no Raman peaks were observed in bilayer samples after the treatment. We conclude that both layers were converted into $WO_{3-x}$. Therefore, different from the RT treatment, more than one layer of $WSe_2$ can be converted by $O_2$ plasma at elevated temperatures.

The surface chemistry alteration of CVD $WSe_2$ flakes upon $O_2$ plasma exposure at different temperatures was investigated by x-ray photoelectron spectroscopy (XPS). Detailed information can be found in the Experimental Section. Figure 2b, c show W 4*f* and Se 3*d* core level spectra of a pristine $WSe_2$ flake and following $O_2$ plasma exposure for 60s at RT, 90 °C, 150 °C, and 250 °C. The dominant XPS signal originated from the substrate ($SiO_2$/Si) due to the lateral size and thickness of $WSe_2$ flakes is shown in Section II (Supporting Information). The XPS scan of pristine $WSe_2$ flake did not reveal any additional chemical states such as W-O or Se-O. In Figure 2b, after $O_2$ plasma exposure at RT, an additional chemical state corresponding to W-O chemical bond was detected at 36.2 eV in W 4*f* core level. No additional chemical states were detected in Se 3*d* core level, indicating that oxygen did not react with selenium. Similar to the $O_2$ plasma treatment at RT, the W-O chemical state was detected in W 4*f* chemical state following the treatment at 90 °C, 150 °C and 250 °C with increasing intensity. Throughout the treatment at different temperatures, the oxide species were below the XPS detection limit in Se 3*d* core level spectra. Figure 2d presents an increase in the percentage of oxidized W atoms on the top few $WSe_2$ layers after $O_2$ plasma treatment at different temperatures. In Figure 2e, the selenium to tungsten ratio of the $WSe_2$ flakes was calculated using the integrated intensity of XPS core levels and corresponding sensitivity factor. It shows that the Se/W ratio decreased gradually from 2.1 in pristine flakes to 0.6 after the treatment at 250 °C, suggesting that the density of W-O chemical states depends critically on the treatment temperature. The W 4*f* and Se 3*d* core level binding energies after each treatment are shown in Figure 2f. The red shift of binding energy was detected in both W4*f* and Se 3*d* core level spectra in all treatment temperatures, indicating different levels of p-doping. Lower binding energy suggests that more prominent doping effect can be achieved at higher temperature treatment.[37,38]

The doping effect in $WSe_2$ was also examined by Raman spectroscopy. Figure 2g compares the Raman spectra of $WSe_2$ flakes before and after the $O_2$ plasma treatment at 250 °C. The degenerate $E^1_{2g}/A_{1g}$ vibrational mode at 250.8 $cm^{-1}$ and the higher wave number peak 2LA(M) at 259.0 $cm^{-1}$ as a Raman fingerprint of $WSe_2$ were detected. A clear blue shift of $E^1_{2g}/A_{1g}$ and 2LA(M) was observed after the 250 °C treatment. According to previous studies[37,39], the blue shift (~ 1.3 $cm^{-1}$) in the $E^1_{2g}/A_{1g}$ peak is correlated to p-type doping effect in $WSe_2$ flakes, which is consistent with our XPS analysis and electrical characterizations discussed in the later paragraph. Therefore, we have confirmed that both atomically precise layer control and doping level modulation can be achieved through different treatment temperatures.

HR-STEM measurements were performed to directly quantify the number of $WSe_2$ layers being converted to oxide by the 250 °C $O_2$ plasma treatment. Detailed information of HR-STEM can be found in the Experimental Section. Figure 3a shows a cross-section view of a $WSe_2$ device with a channel length ($L_G$) of 65nm. The observed bending curvature was caused by the carbon layer deposition induced stress during the TEM sample preparation using focused ion beam (FIB). From the magnified image of the channel region presented in Figure 3b, nearly three layers of $WSe_2$ were converted to $WO_{3-x}$ by the 250 °C treatment, leaving two $WSe_2$ layers remaining underneath. Interestingly, it was observed that $WO_{3-x}$ penetrated laterally into the contact at the scale of ~12nm, as revealed in Figure 3c. We believe this phenomenon contributes significantly to unpinning of the metal Fermi level and potentially lowering of the SB height for easier hole injection. Furthermore, $WO_{3-x}$ induced heavy doping in the remaining $WSe_2$ under the contact can effectively reduce the depletion width to produce a transparent barrier for carrier injection into the channel. Both mechanisms contribute to a very low contact resistance of 528 Ω μm and correspondingly record high on-state performance reported in the later paragraph. More detailed information regarding the EELS mapping can be found in Section III (Supporting Information).

## 2.2. Electrical Characterizations

We now focus on electrical characterization of devices that have undergone $O_2$ plasma treatment at different temperatures, as shown in Figure 4a. One could immediately observe significant differences in the magnitude of the threshold voltage ($V_{TH}$) shift from the pristine (black) to after treatment (red) characteristics. It is clear that $V_{TH}$ shift, an indication of the doping level, increases with increasing temperature, consistent with the shift of binding energy shown in Figure 2f. We believe that the higher doping level achieved at higher temperature can be attributed to a larger number of $WSe_2$ layers being converted into a thicker $WO_{3-x}$ layer. Except for the device treated at 250 °C, $V_{TH}$ of the other two devices treated at RT and 150 °C is within the voltage window to reveal the off-state performance. The preserved on/off ratios of ~$10^7$ indicate that $WO_{3-x}$ serves as an effective doping layer rather than a conductive layer shunting between the S/D electrodes.[40] Based on the $V_{TH}$ shift extracted from the device characteristics, we can calculate the amount of charges induced by doping ($Q = C_{ox}V_{TH_Shift}$, $C_{ox}$ = 38.5 nF/cm$^2$) and estimate the dopant concentration to be ~ 2.2x$10^{12}$ cm$^{-2}$ for devices undergone the RT treatment and ~ 8.3 $10^{12}$ cm$^{-2}$ for those gone through the treatment at 150 $^oC$. More devices data set can be found in Section IV (Supporting Information)

In addition, device contact resistances ($R_C$) were significantly reduced after the plasma treatment. 4-probe configuration shown in Figure 1b was used to extract $R_C$. Figure 4b, c show $R_C$ as a function of back gate voltage for pristine devices and the same set of devices after the $O_2$ plasma treatment at 150 °C and 250 °C, respectively. Due to the positive $V_{TH}$ shift, $R_C$ is less gate voltage dependent and reduced drastically. $R_C$ of a device gone through the 250 $^oC$ treatment was extracted to be as low as 528 Ω μm, at $V_{BG}$ = -50V. As explained earlier, we believe that the $O_2$ plasma treatment not only affects the channel doping but also lowers the SB height and barrier width at the contact interface to allow for higher current injection, which is now attributed to the lateral penetration of $WO_{3-x}$ as observed from our HR-STEM analysis. A table of extracted $R_C$ values from different devices with or without treatment can be found in Section V (Supporting Information).

Utilizing the demonstrated doping and low $R_C$, we fabricated devices with scaled channel length ($L_G$) and achieved the outstanding on-state performance in $WSe_2$ FETs. Twelve SB-devices with different $L_G$ ranging from ~70 nm to ~1050 nm were fabricated on exfoliated multilayer (5-10 layers) $WSe_2$. 250 $^oC$ $O_2$ plasma treatment was performed to all devices. Total device resistance ($R_{total}$) and current density ($I_{DS}$) extracted at $V_B$ = -50V and $V_{DS}$ = -0.9V are shown in Figure 5a. Although these devices were not fabricated on the same flake to guarantee an accurate extraction of $R_C$ from the transmission line method (TLM), we still performed the extraction to get a rough estimate. $R_C$ ~ 1.1 kΩ μm was extracted from the linear fitting of the $R_{total}$ vs $L_G$ plot, which agrees with the values obtained from the 4-probe measurements, presented in Table S1 (Supporting Information). The output characteristics of our best performing device with $L_G$ ~ 70nm is shown in Figure 5b with $I_{DS}$ = 320 μA/μm being achieved at $V_{DS}$ = -1V. Finally, these $O_2$ plasma treated devices were placed in the laboratory ambient environment without any passivation layer for 7 days before re-measurements. Negligible changes in characteristics were observed as shown in Section VI (Supporting Information), suggesting a robust p-doping scheme for high performance $WSe_2$ devices.

Finally, we compare our result to other reported contact resistance for hole injection in $WSe_2$-based devices,[22,26–28,41–44] and summarize in Table 1. Our $O_2$ plasma treatment offers a comparable $R_C$ while requiring a simpler fabrication process compared to a 2D/2D contact.[22]

## 3. Conclusion

We have achieved tunable p-type doping on $WSe_2$ through $O_2$-plasma treatment at different temperatures, with supporting evidences from XPS, Raman, HR-STEM and electrical characteristics. We conclude that the doping level is determined by how many $WSe_2$ layers being converted into $WO_{3-x}$ by the $O_2$-plasma treatment. Larger number of layers are converted at higher temperatures, resulting in thicker $WO_{3-x}$ for higher doping. The penetration of $WO_{3-x}$ into the contact region is believed to contribute to the unpinning of the Fermi level and thinning of the barrier width for higher current injection. Low $R_C$ ~528 Ω μm was measured from 4-probe measurements after 250 $^oC$ $O_2$ plasma treatment, leading to a record-high hole current in $WSe_2$ devices. This air-stable, efficient p-doping strategy can enable high performance $WSe_2$-based electronics or be applied to other material of interests for tunable doping effect by transferring $WSe_2$ on top followed by self-limiting oxidation under specific temperature treatment.

## 4. Methods

**CVD Flakes Transferring Process:** Mono/bi/tri-layer CVD $WSe_2$ flakes were purchased from 2D Layer (https://2dlayer.com/). To transfer onto a desired substrate, polystrene (PS) was used as the supporting film to peel off the $WSe_2$ flakes from the growth substrate. 9 g of PS (Molecular weight ~192 000 g/mol) was dispersed in 50 mL toluene. Then this solution was spin-coated on the growth substrate at a speed of 4000 rpm for 40 secs, followed by baking at 90 $^oC$ for 5 mins. In order to allow water to penetrate to the interface between the $WSe_2$ film and the $SiO_2$/Si substrate to detach the $WSe_2$ flakes, a diamond scribe was used to make some scratches at the edges of the PS film. Next, the PS film attached to the $WSe_2$ flakes was gradually peeled off from the growth substrate in water and transferred to the target substrate. Finally, the PS film was removed by soaking in toluene, acetone and IPA.

**2-probe and 4-probe Device Fabrication:** Mono/bi-layer CVD $WSe_2$ flakes were transferred or multilayer layers (5-10) $WSe_2$ was exfoliated from a bulk crystal onto a 90nm $SiO_2$ capped $p^{++}$ doped Si substrate as a global back-gated scheme. E-beam lithography was employed to define source/drain (S/D) regions (for 2-probe devices) and two additional voltage probes (for 4-probe devices) followed by e-beam evaporated Ti (1nm) / Pd (30nm) (at pressure ~ 1E-6 torr) as electrodes and a PMMA lift-off process.

**Electrical Characterization:** The electrical measurements were performed by HP 4156B precision semiconductor parameter analyzer with Lake Shore probe station under vacuum at room temperature.

**X-ray Photoelectron Spectroscopy:** XPS scans were carried out in an Ultra High Vacuum (UHV) cluster tool using an Omicron EA125 hemispherical 7 channels analyzer with a monochromatic Al Kα source (hν= 1486.7 eV).[45] XPS peaks were deconvoluted and analyzed using Aanalyzer software.[46] Quantitative analysis of the elemental concentration of the samples is acquired based on integrated photoelectron intensity and sensitivity factor for a given core level.

Quantitative analysis of relative elemental concentration can be determined from XPS measurement. The number of photoelectrons per second for specific elemental core level (I) is directly proportional to the number of atoms of the elements per centimeter cubic of the sample surface (n), while it is indirectly proportional to atomic sensitivity factor (S). Atomic sensitivity factor (S) of the elements core level is developed from a specific spectrometer. Thus, the elemental concentration ($C_x$) is described as: $C_x = \frac{n_x}{n_i} = \frac{I_x/S_x}{\sum I_i/S_i}$.

**Raman Spectroscopy:** Raman spectra were taken using a 532 nm laser focused through a 100X objective lens at room temperature and under ambient condition.

**TEM Analysis:** TEM analysis is carried out using an aberration-corrected (probe Cs-corrector) JEM-ARM200F (JEOL. USA. Inc.) electron microscope operated at 200 kV. The high angle annular dark field (HAADF) and annular bright field (ABF) images are performed to study the cross-sectional morphology of the device. The convergence semi-angle of the electron probe is set to 25 mrad, and the collection semi-angle is 70-250 mrad for HAADF and 12-24 mrad for ABF, respectively. The elemental characterization of the device is performed using the electron energy loss spectroscopy (EELS) line scan using a Gatan Enfina spectrometer with the collection semi-angle for EELS for 30 mrad.

## References

1. Jena, D. Tunneling transistors based on graphene and 2-D Crystals. *Proc. IEEE* **101**, 1585 (2013).
2. Seabaugh, B. A. C. &Zhang, Q. Low-Voltage Tunnel Transistors for Beyond CMOS Logic. *Proc. IEEE* **98**, 2095 (2010).
3. Müller, M. R. *et al.* Gate-Controlled WSe2 Transistors Using a Buried Triple-Gate Structure. *Nanoscale Res. Lett.* **11**, 512 (2016).
4. Pang, C.-S., Ilatikhameneh, H. &Chen, Z. Gate tunable 2D WSe2 Esaki diode by SiNx doping. in *Device Research Conference - Conference Digest, DRC* 1–2 (2017). doi:10.1109/DRC.2017.7999450
5. Sarkar, D. *et al.* A subthermionic tunnel field-effect transistor with an atomically thin channel. *Nature* **526**, 91 (2015).
6. Pang, C.-S. *et al.* WSe 2 Homojunction Devices : Electrostatically Configurable as Diodes , MOSFETs , and Tunnel FETs for Reconfigurable Computing. *Small* 1902770 (2019). doi:10.1002/smll.201902770
7. Pang, C.-S., Thakuria, N., Gupta, S. K. &Chen, Z. First Demonstration of WSe 2 Based CMOS-SRAM. in *IEEE Int. Electron Devices Meet.* 22.2.1-22.2.4 (IEEE, 2018). doi:10.4231/D3ZC7RV9X
8. Pang, C.-S. &Chen, Z. First Demonstration of WSe 2 CMOS Inverter with Modulable Noise Margin by Electrostatic Doping. in *Device Research Conference - Conference Digest, DRC* 1–2 (IEEE, 2018). doi:10.1109/DRC.2018.8442258
9. Paul, T., Ahmed, T., Tiwari, K. K. &Thakur, C. S. A high-performance MoS 2 synaptic device with floating gate engineering for neuromorphic computing A high-performance MoS 2 synaptic device with floating gate engineering for neuromorphic computing. *2D Mater.* **6**, 045008 (2019).

10. Seo, S. *et al.* Artificial optic-neural synapse for colored and color-mixed pattern recognition. *Nat. Commun.* **9**, 5106 (2018).

11. Xie, D., Hu, W. &Jiang, J. Bidirectionally-trigged 2D MoS 2 synapse through coplanar-gate electric- double-layer polymer coupling for neuromorphic complementary spatiotemporal learning. *Org. Electron.* **63**, 120 (2018).

12. Gao, L. Flexible Device Applications of 2D Semiconductors. *Small* **13**, 1603994 (2017).

13. Lin, P. *et al.* Piezo-Phototronic Effect for Enhanced Flexible MoS 2 / WSe 2 van der Waals Photodiodes. *Adv. Funct. Mater.* **28**, 1802849 (2018).

14. Yu, X., Prévot, M. S., Guijarro, N. &Sivula, K. Self-assembled 2D WSe 2 thin films for photoelectrochemical hydrogen production. *Nat. Commun.* **6**, 7596 (2015).

15. Akinwande, D., Petrone, N. &Hone, J. Two-dimensional flexible nanoelectronics. *Nat. Commun.* **5**, 5678 (2015).

16. Desai, S. B. *et al.* MoS 2 transistors with 1-nanometer gate lengths. *Science (80-. ).* **354**, 99 (2016).

17. Nourbakhsh, A. *et al.* MoS 2 Field-Effect Transistor with Sub-10 nm Channel Length. *Nano Lett.* **16**, 7798 (2016).

18. Cui, X. *et al.* Multi-terminal transport measurements of MoS2 using a van der Waals heterostructure device platform. *Nat. Nanotechnol.* **10**, 534–540 (2015).

19. Leong, W. S. *et al.* Low Resistance Metal Contacts to MoS2 Devices with Nickel-Etched-Graphene Electrodes. *ACS Nano* 869–877 (2015). doi:10.1021/nn506567r

20. Smithe, K. K. H., English, C. D., Suryavanshi, S.V &Pop, E. Intrinsic electrical transport and performance projections of synthetic monolayer MoS 2 devices Intrinsic electrical transport and performance projections of synthetic monolayer MoS 2 devices. *2D Mater.* **4**, 011009 (2017).

21. English, C. D., Shine, G., Dorgan, V. E., Saraswat, K. C. &Pop, E. Improved Contacts to MoS 2 Transistors by Ultra-High Vacuum Metal Deposition. *Nano Lett.* **16**, 3824 (2016).

22. Chuang, H. J. *et al.* Low-Resistance 2D/2D Ohmic Contacts: A Universal Approach to High-Performance WSe 2 , MoS 2 , and MoSe 2 Transistors. *Nano Lett.* **16**, 1896–1902 (2016).

23. Kim, C. *et al.* Fermi Level Pinning at Electrical Metal Contacts of Monolayer Molybdenum Dichalcogenides. *ACS Nano* **11**, 1588 (2017).

24. Gong, C., Colombo, L., Wallace, R. M. &Cho, K. The Unusual Mechanism of Partial Fermi Level Pinning at Metal − MoS2 Interfaces. *Nano Lett.* **14**, 1714 (2014).

25. Bampoulis, P. *et al.* Defect Dominated Charge Transport and Fermi Level Pinning in MoS2 / Metal Contacts. *ACS Appl. Mater. Interfaces* **9**, 19278 (2017).

26. Yamamoto, M., Nakaharai, S., Ueno, K. &Tsukagoshi, K. Self-Limiting Oxides on WSe 2 as Controlled Surface Acceptors and Low-Resistance Hole Contacts. *Nano Lett.* **16**, 2720–2727 (2016).

27. Cai, L. *et al.* Rapid flame synthesis of atomically thin MoO3 down to monolayer thickness for effective hole doping of WSe2. *Nano Lett.* **17**, 3854 (2017).

28. Fang, H. *et al.* High-Performance Single Layered WSe 2 p-FETs with Chemically Doped Contacts. *Nano Lett.* **12**, 3788 (2012).

29. Chen, K. *et al.* Air stable n-doping of WSe2 by silicon nitride thin films with tunable fixed charge density. *Apl Mater.* **2**, 092504 (2014).

30. Hung, T. Y. T., Pang, C.-S., Liu, X., Zemlyanov, D. &Chen, Z. Atomically Thin p-doping Layer and Record High Hole Current on WSe2. in *Device Research Conference - Conference Digest, DRC* 1–2 (2019).

31. Smyth, C. M. *et al.* Engineering the interface chemistry for scandium electron contacts in WSe 2 transistors and diodes Engineering the interface chemistry for scandium electron contacts in WSe 2 transistors and diodes. *2D Mater.* **6**, 045020 (2019).

32. Pudasaini, P. R. *et al.* High-performance multilayer WSe2field-effect transistors with carrier type control. *Nano Res.* **11**, 722 (2018).

33. Smyth, C. M. *et al.* Engineering the Palladium − WSe 2 Interface Chemistry for Field E ff ect Transistors with High-Performance Hole Contacts. *ACS Appl. Nano Mater.* **2**, 75 (2019).

34. Das, S. &Appenzeller, J. WSe2 field effect transistors with enhanced ambipolar characteristics. *Appl. Phys. Lett.* **103**, 103501 (2013).

35. Luo, X. *et al.* Effects of lower symmetry and dimensionality on Raman spectra in two-dimensional WSe 2. *Phys. Rev. B* **88**, 195313 (2013).

36. Li, Z. *et al.* Layer Control of WSe2 via Selective Surface Layer Oxidation. *ACS Nano* **10**, 6836–6842 (2016).

37. Khosravi, A. *et al.* Covalent nitrogen doping in molecular beam epitaxy-grown and bulk WSe2. *APL Mater.* **6**, 026603 (2018).

38. Addou, R. *et al.* One dimensional metallic edges in atomically thin WSe 2 induced by air exposure. *2D Mater.* **5**, 025017 (2018).

39. Desai, S. B., Seol, G., Kang, J. S., Fang, H. &Battaglia, C. Strain-Induced Indirect to Direct Bandgap Transition in Multilayer. *Nano Lett.* **14**, 4592 (2014).

40. Liu, B. *et al.* High-Performance WSe 2 Field-E ff ect Transistors via Controlled Formation of In- Plane Heterojunctions. *ACS Nano* **10**, 5153 (2016).

41. Jung, Y. *et al.* Transferred via contacts as a platform for ideal two-dimensional transistors. *Nat. Electron.* **2**, 187–194 (2019).

42. Chien, P. Y. *et al.* Reliable doping technique for WSe2 by W:Ta co-sputtering process. in *2016 IEEE Silicon Nanoelectronics Workshop, SNW 2016* 58–59 (2016). doi:10.1109/SNW.2016.7577984

43. Zhao, P. *et al.* Air stable p-doping of WSe2 by covalent functionalization. *ACS Nano* **8**, 10808–10814 (2014).

44. Zhang, R., Drysdale, D., Koutsos, V. &Cheung, R. Controlled Layer Thinning and p-Type Doping of WSe2 by Vapor XeF2. *Adv. Funct. Mater.* **27**, 1702455 (2017).

45. R.M.Wallace. In-Situ Studies of Interfacial Bonding of High-N Dielectrics for CMOS Beyond 22nm. *ECS Trans.* **16**, 255 (2008).

46. Herrera-Go´mez, A., Hegedus, A. &Meissner, P. L. Chemical depth profile of ultrathin nitrided films. *Appl. Phys. Lett.* **81**, 1014 (2002).

**Acknowledgements**

C.-S. P. And T.Y.T. H. contributed equally to this work.

C.-S. P., T.Y.T. H., A. K., R. A., R. M. W., and Z. C. acknowledge financial support from NEWLIMITS, a center in nCORE, a Semiconductor Research Corporation (SRC) program sponsored by NIST through award number 70NANB17H041. M. J. K. was supported in part by the Louis Beecherl, Jr. Endowment Funds, Global Research and Development Center Program (2018K1A4A3A01064272) and Brain Pool Program (2019H1D3A2A01061938) through the National Research Foundation of Korea (NRF) funded by the Ministry of Science and ICT.

**Author Contributions**

Z. C. conceived and managed the project. C.-S. P and T. H. fabricated the devices, perform the $O_2$-plasma treatment, and conducted the electrical and Raman measurements. A. K. and R. A.

performed the XPS measurement and analysis. Q. W. performed the FIB on samples for HRTEM analysis. All authors wrote and revised the manuscript.

**Competing Interests Statement**

The authors declare no competing interests.

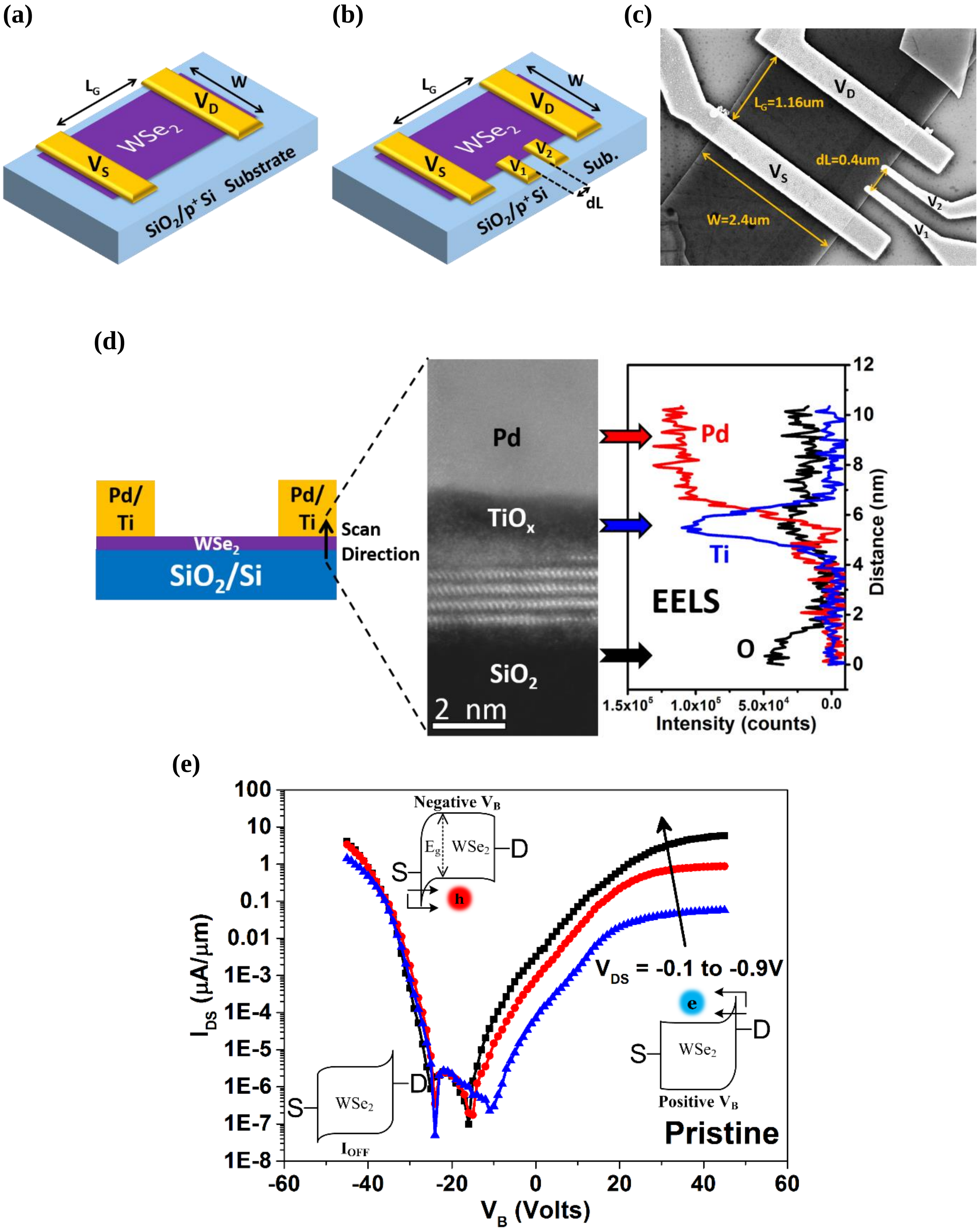

**Figure 1.** Schematics illustrations of a) 2-probe and b) 4-probe (type 2) $WSe_2$ devices. c) SEM image of a 4-probe device with dimensions labelled. d) HR-STEM and EELS line scan across the $SiO_2$/$WSe_2$/contact region. e) Transfer characteristics of a $WSe_2$ SB device showing ambipolar behaviors with gate-dependent electron/hole injection.

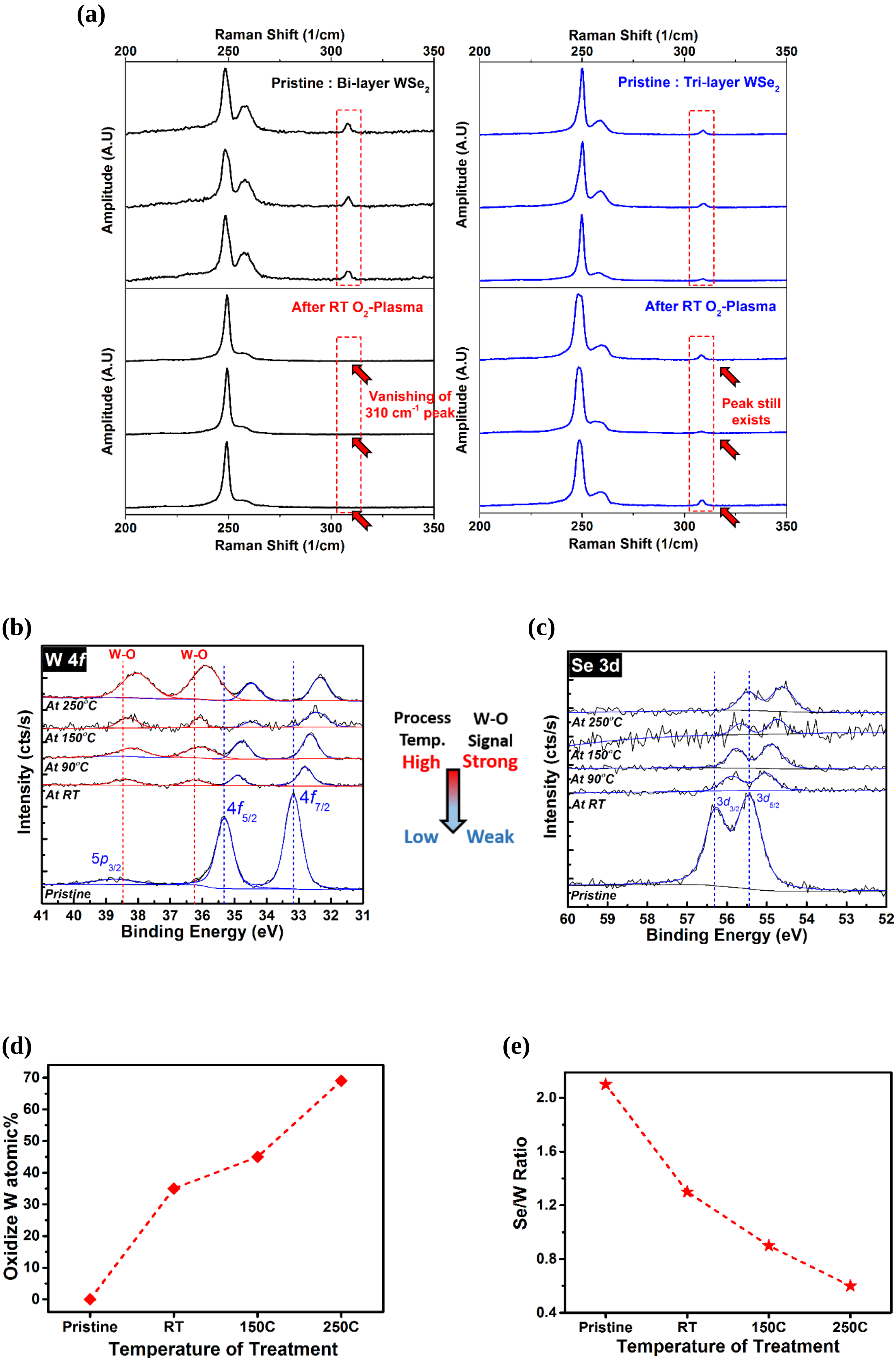

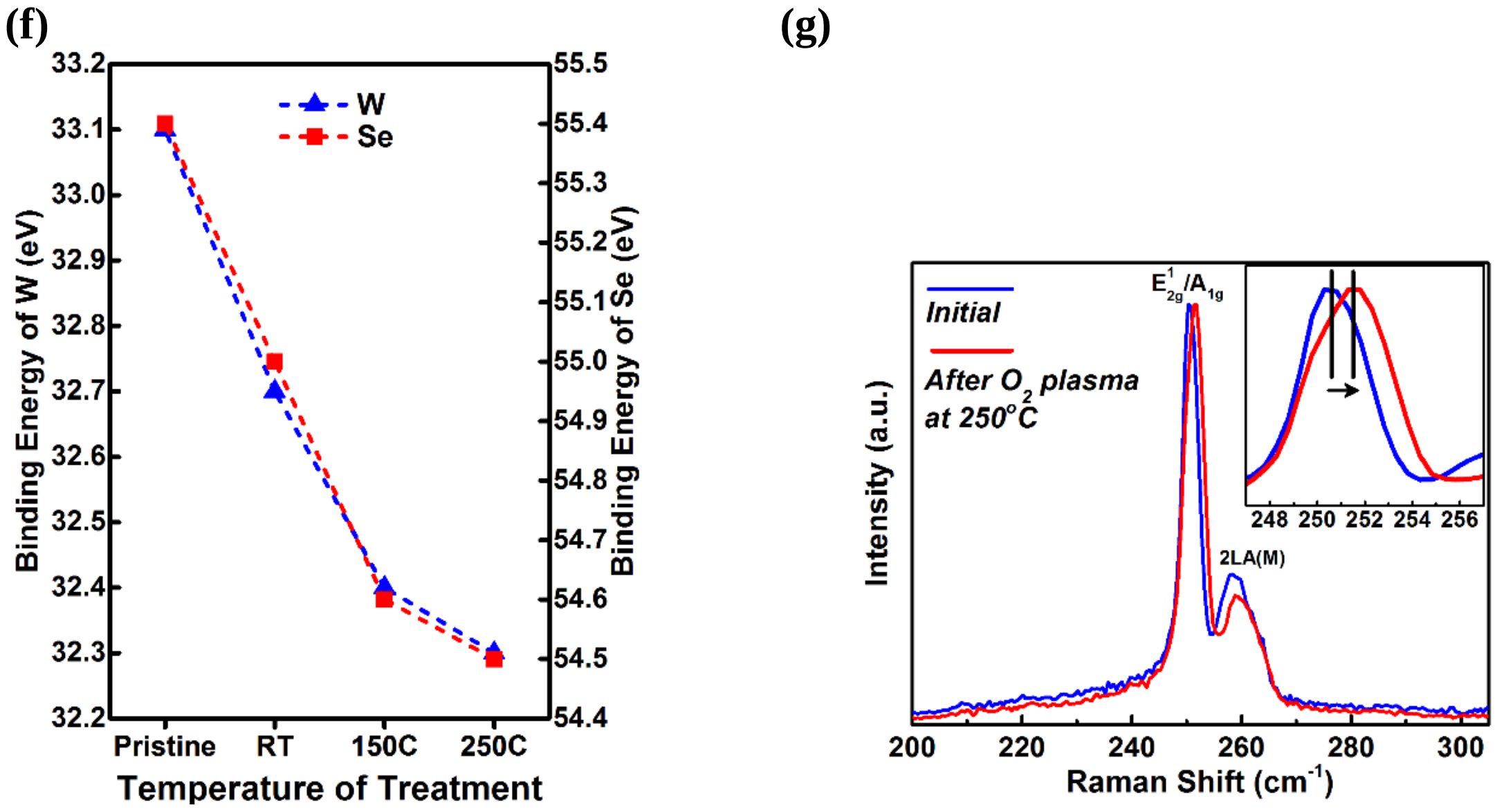

**Figure 2.** a) Comparison of Raman spectra between pristine bi-layer/tri-layer $WSe_2$ and after RT $O_2$ plasma treatment. b) W 4f and c) Se 3d core level spectra of pristine $WSe_2$ and after $O_2$ plasma treatment at RT, 90 $^oC$, 150 $^oC$, or 250 $^oC$. d) Percentage of oxidized W atoms and e) Se to W ratio on the top few layers in pristine $WSe_2$ and after $O_2$ plasma treatment at various temperatures. f) The red shift of the binding energy calculated from b, c after $O_2$ plasma treatment at different temperatures. g) Raman spectra of $WSe_2$ before and after treatment at 250 $^oC$. The blue shift of $E^1_{2g}/A_{1g}$ and 2LA(M) indicate p-doping effect in $WSe_2$.

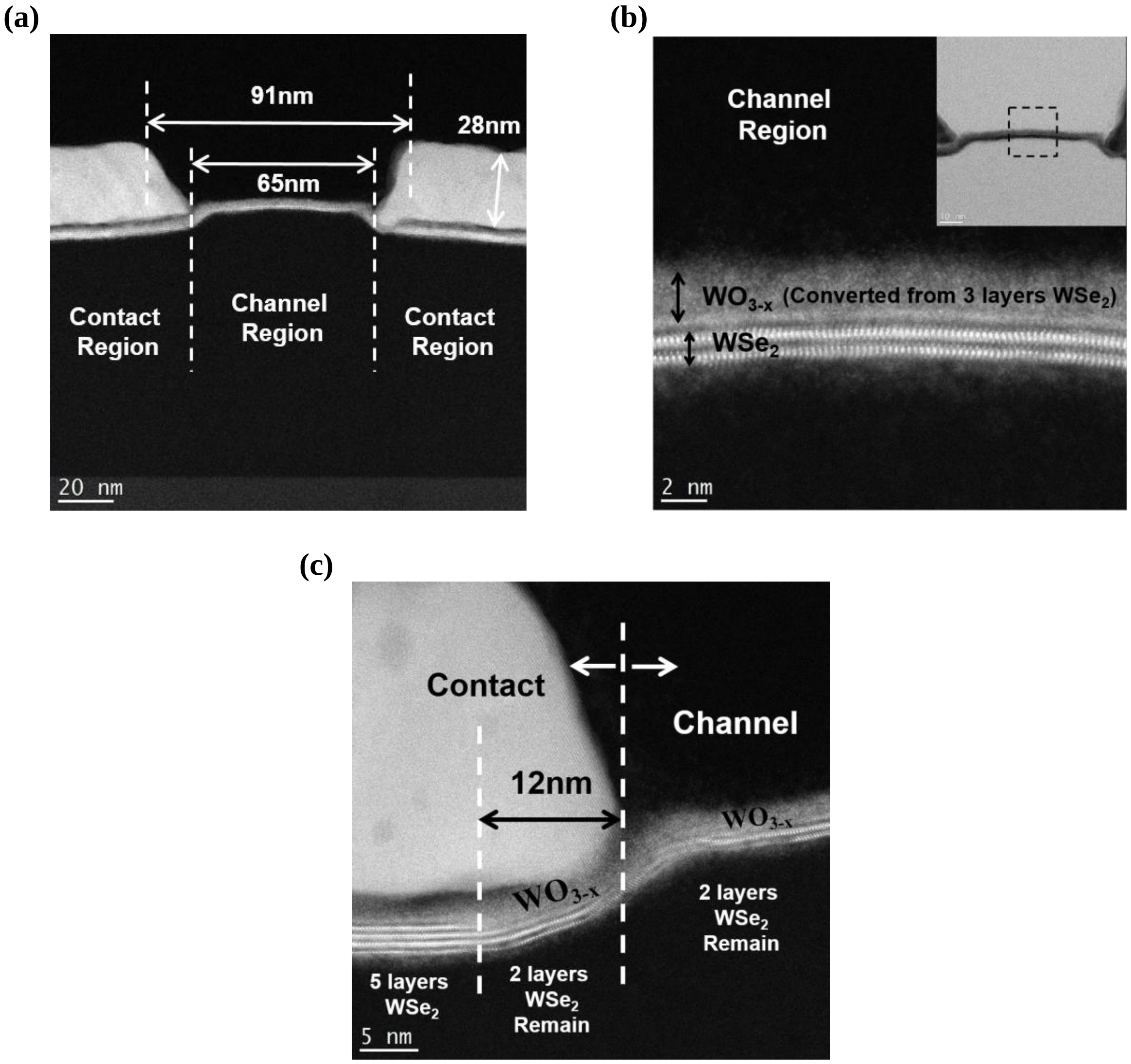

**Figure 3.** a) Cross-section view of a $WSe_2$ device with $L_G$ = 65nm. b) The zoom-in observation of the channel region from a, indicating ~ 3 layers of $WSe_2$ is converted into $WO_{3-x}$. c) The lateral penetration of $WO_{3-x}$ underneath the edge of contact after the $O_2$-plasma treatment.

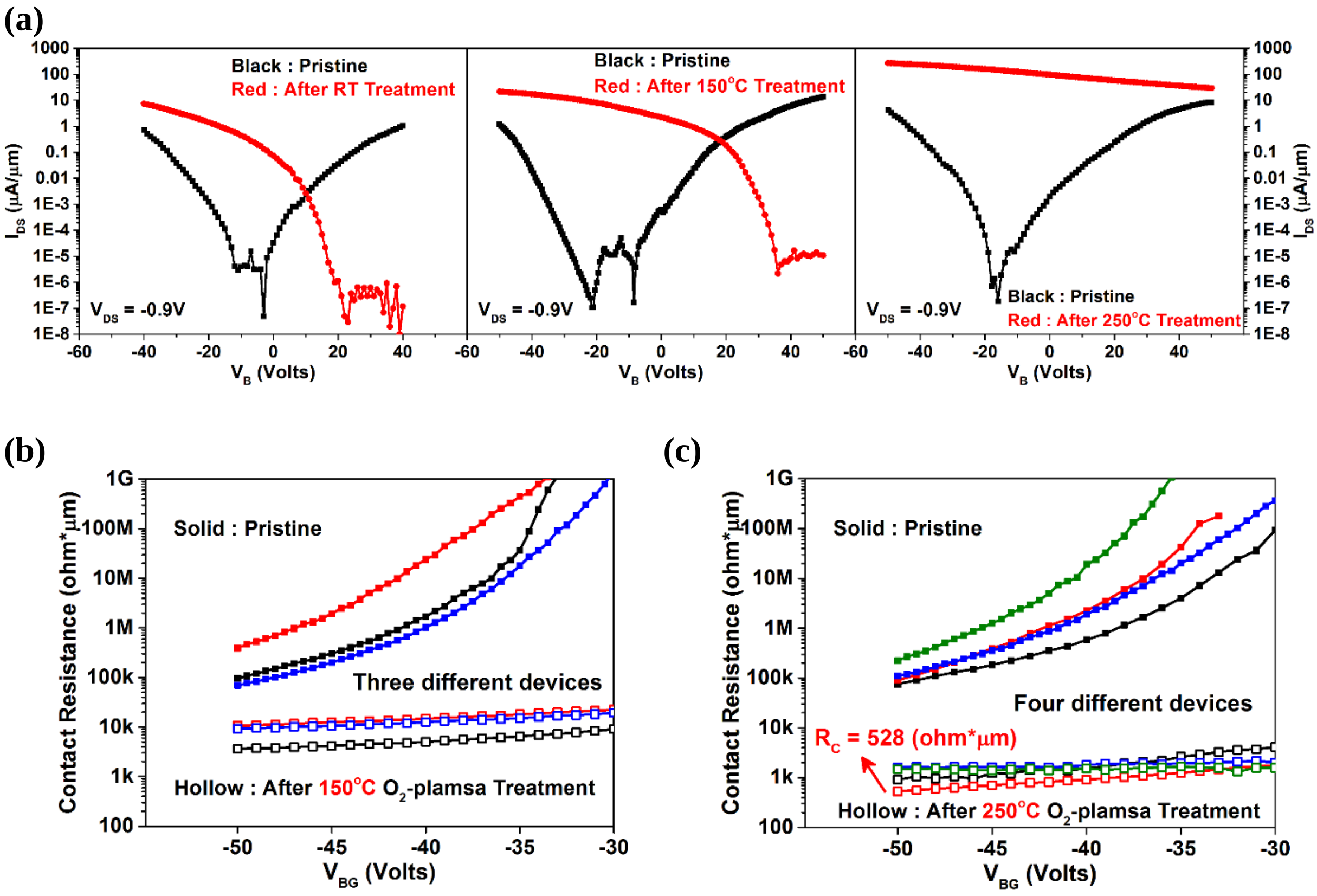

**Figure 4.** a) Transfer characteristics comparison of pristine $WSe_2$ FETs and after $O_2$ plasma treatment at RT, 150 $^oC$, and 250 $^oC$. The comparison of $R_C$ between pristine and after $O_2$ plasma treatment at b) 150 $^oC$ and c) 250 $^oC$ using 4-probe measurements.

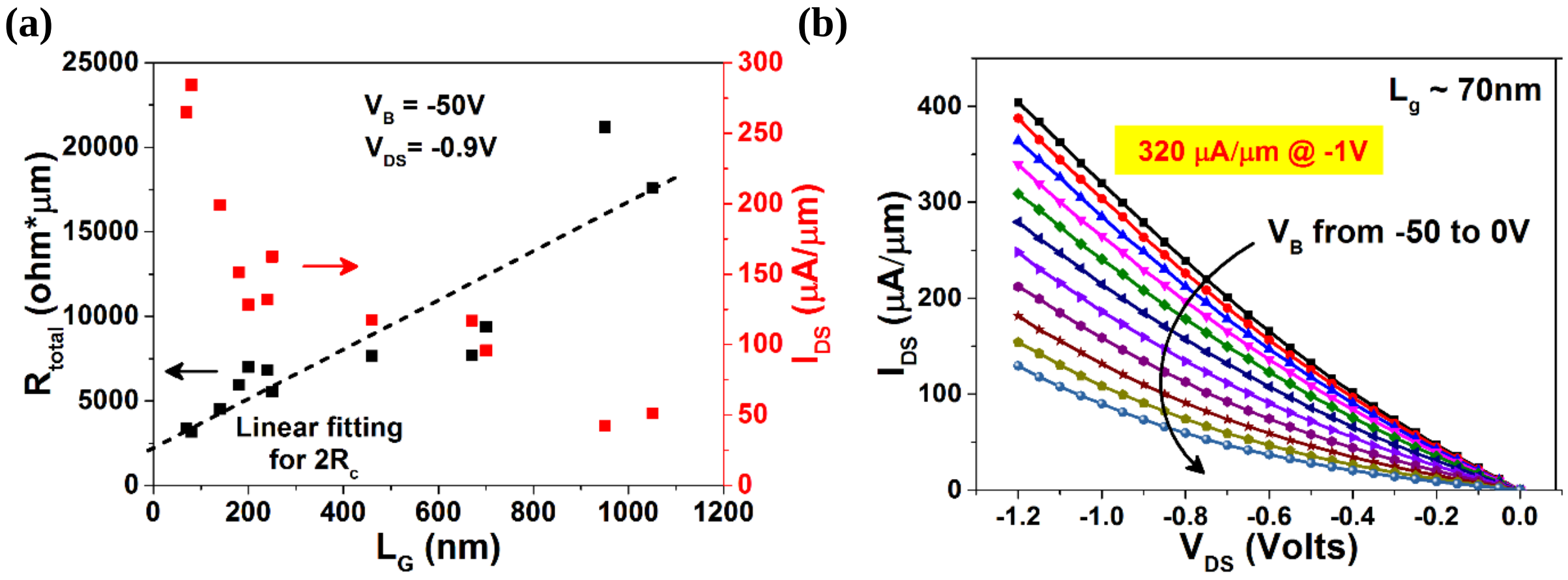

**Figure 5.** a) Measured $R_{total}$ (black data) and current density (red data) at $V_B$ = -50V and $V_{DS}$ = -0.9V for 12 devices with $L_G$ ranging from ~70nm to ~1050nm. $R_C$ ~ 1.14 kΩ µm is extracted from a linear fitting curve. b) Output characteristics of a device with record-high hole current density of 320 µA/um at $V_{DS}$ = -1V after 250 $^oC$ $O_2$ plasma treatment.

**Table 1.** $R_C$ comparison for $WSe_2$ hole injection at room temperature.

| **Layer Number [ref]** | **$R_C$ (kΩ*µm)** | **Special Treatment** |
|---|---|---|
| 10L [22] | 0.3 | 2D/2D Contact |
| 6L [27] | 0.9 | $MoO_3$ Passivation |
| 4L [26] | 1.1 | Ozone Treatment |
| 10L [28] | 1.3 | $NO_2$ Treatment |
| 2L [41] | 4.0 | Transferred Contact |
| 20L [42] | 11.4 | W:Ta Co-sputtering |
| 13L [41] | 12.5 | Transferred Contact |
| 1L [43] | 38.0 | $NO_2$ Treatment |
| 1L [41] | 50.0 | Transferred Contact |
| 7L [44] | 100 | $XeF_2$ Thinning |
| **~8L (this work)** | **0.5** | **$O_2$-plasma at 250 $^{o}$*C*** |